\documentclass{amsart}
\usepackage[top=0.9in, bottom=0.85in, left=0.9in, right=0.9in]{geometry}

\usepackage{amsmath,amsthm,amssymb}
\usepackage{txfonts}

\theoremstyle{plain}
\newtheorem{theorem}{Theorem}[section]

\theoremstyle{definition}
\newtheorem{definition}{Definition}[section]
\newtheorem{example}{Example}[section]

\numberwithin{equation}{section}

\begin{document}

\title[The $z$-Transform and Nonhomogeneous Linear Recurrence Equations over Semirings]{The $z$-Transform and Automata-Recognizable Systems of Nonhomogeneous Linear Recurrence Equations over Semirings}
\author[E. Carta-Gerardino]{Edoardo Carta-Gerardino \\ Department of Mathematics and Computer Science \\ York College $|$ City University of New York \\ Jamaica, NY 11451 \\ \\ ecarta-gerardino@york.cuny.edu}





\begin{abstract}
A nonhomogeneous system of linear recurrence equations can be recognized by an automaton $\mathcal{A}$ over a one-letter alphabet $A = \{z\}$. Conversely, the automaton $\mathcal{A}$ generates precisely this nonhomogeneous system of linear recurrence equations. We present the solutions of these systems and apply the $z$-transform to these solutions to obtain their series representation. Finally, we show some results that simplify the series representation of the $z$-transform of these solutions. We consider single systems as well as the composition of two systems.

{\bf Keywords}: $z$-transform, automata, nonhomogeneous linear recurrence equations, semirings
\end{abstract}

\maketitle

\vspace{-0.5cm}

\section{Introduction}

Consider a finite automaton $\mathcal{A}$ over a one-letter alphabet $A = \{z\}$, with weights in a semiring $R$. Since $A = \{z\}$, an arbitrary word $w \in A^*$ has the form $w = z \cdot z \cdots z = z^n$, for some $n \in \mathbb{N}$. By definition, the weight of a word $w = z^n$ equals the sum of the weights of all paths with label $w = z^n$ \cite{Dro05,Rut02}. But since every transition reads the letter $z$, the weight of the word $z^n$ equals the sum of the weights of all paths of length $n$. Note that, in this kind of automaton, we are not interested in accepting/rejecting words over some alphabet, but rather in counting paths of length $n$, and keeping track of the weights of such paths. The idea of using automata as counting devices has been used recently with applications in combinatorics and difference equations \cite{Rue01,Rut01,Rue02}, but goes back to classical automata theory \cite{Sch62}.

Now suppose we are interested in computing the weight of all paths of length $n$ (equivalently, all paths with label $z^n$) in an automaton $\mathcal{A}$, starting at a specific state $p$. Then we would look at all paths starting at $p$, compute the weight of every path of length $n$, and add up the results. Denote this function by $f_p$, and notice that for every $n \in \mathbb{N}$, $f_p(n)$ gives the sum of the weights of all paths of length $n$ starting at state $p$. Hence, in this kind of automaton, every state $p$ generates a function $f_p \in R^{\mathbb{N}}$.

Let $\mathcal{A}$ be a weighted $k$-state automaton over $A= \{z\}$ and an arbitrary semiring $R$, and let $f_1,f_2,\ldots,f_k \in R^{\mathbb{N}}$ be the functions generated by states $p_1,p_2,\ldots,p_k$, respectively. Then it can be shown \cite{Car08,Rut01} that these functions satisfy the following system of homogeneous {\bf linear recurrence equations} with constant coefficients:

\begin{gather*}
f_1(n+1) = a_{11}f_1(n) + a_{12}f_2(n) + \ldots + a_{1k}f_k(n) \\
f_2(n+1) = a_{21}f_1(n) + a_{22}f_2(n) + \ldots + a_{2k}f_k(n) \\
\vdots \\
f_k(n+1) = a_{k1}f_1(n) + a_{k2}f_2(n) + \ldots + a_{kk}f_k(n)
\end{gather*}

\noindent where $a_{ij} \in R$ is the weight of the transition $p_i \longrightarrow p_j$. Conversely, given this system of linear recurrence equations, the weighted finite automaton recognizing it is precisely $\mathcal{A}$.

This result can be generalized to consider the case when the system is {\it nonhomogeneous} and  has {\it variable coefficients} \cite{Car08}. That is, there is an automaton $\mathcal{A}$ over $A= \{z\}$ and $R$ that generates the following system of linear recurrence equations:

\begin{gather*}
f_1(n+1) = a_{11}(n)f_1(n) + a_{12}(n)f_2(n) + \ldots + a_{1k}(n)f_k(n) + g_1(n) \\
f_2(n+1) = a_{21}(n)f_1(n) + a_{22}(n)f_2(n) + \ldots + a_{2k}(n)f_k(n) + g_2(n) \\
\vdots \\
f_k(n+1) = a_{k1}(n)f_1(n) + a_{k2}(n)f_2(n) + \ldots + a_{kk}(n)f_k(n) + g_k(n)
\end{gather*}

\noindent where $f_i,g_i,a_{ij} \in R^{\mathbb{N}}$. And conversely, given this system of linear recurrence equations, the automaton recognizing it is precisely $\mathcal{A}$. The structure of this automaton is different from the structure of the automaton recognizing the homogeneous system. The difference accounts for the fact that the nonhomogeneous system contains two types of functions; the functions $g_i$ are the \textit{input} whereas the functions $f_i$ are the \textit{output}. A state in $\mathcal{A}$ stores the current value of the function it represents. If we identify a state with the function it generates, then $a_{ij} \in R^{\mathbb{N}}$ is the variable weight of the transition $f_i \longleftarrow f_j$ and, for every $i$, there is an additional transition $g_i \longrightarrow f_i$, with weight $1 \in R$. This kind of automaton generalizes the one recognizing the homogeneous system as it recognizes the nonhomogeneous system, as well as the homogeneous one.

Let $f(n) = \begin{pmatrix} f_1(n) \\ f_2(n) \\ \vdots \\ f_k(n) \end{pmatrix}$, $g(n) = \begin{pmatrix} g_1(n) \\ g_2(n) \\ \vdots \\ g_k(n) \end{pmatrix}$, and $A(n) = \begin{pmatrix} a_{11}(n) & a_{12}(n) & \cdots & a_{1k}(n) \\ a_{21}(n) & a_{22}(n) & \cdots & a_{2k}(n) \\ \vdots  & \vdots  & \ddots & \vdots \\ a_{k1}(n) & a_{k2}(n) & \cdots & a_{kk}(n) \\ \end{pmatrix}$. Then we can rewrite the above system as $f(n+1) = A(n)f(n) + g(n)$, where $f,g \in R^{\mathbb{N}}_{k \times 1}$ and $A \in R^{\mathbb{N}}_{k \times k}$. Sometimes we consider the (simpler) case when the system has constant coefficients. In that case, we are dealing with the system $f(n+1) = Af(n) + g(n)$, where $f,g \in R^{\mathbb{N}}_{k \times 1}$ and $A \in R_{k \times k}$.

We will first consider the solutions of these systems, and then move on to consider the solutions of composing two systems. We will then apply the {\bf $z$-transform} to these solutions to obtain their series representation. Finally, we will show a result that presents a simplified series representation of their $z$-transform.

\section{Solutions of Nonhomogeneous Systems of Linear Recurrence Equations}

From now on, assume for simplicity that the initial conditions of the systems above are 0. That is, we assume that $f(0)=0$. Then it can be shown \cite{Car08} that the solution of the system with constant coefficients is given by

\begin{equation} \label{eq:2Sec1}
f(n+1) = A^n \ast g(n),
\end{equation}

\noindent where $\ast$ denotes discrete convolution: for functions $\alpha(n)$ and $\beta(n)$, $ (\alpha \ast \beta)(n) = \sum_{i=0}^n \alpha(n-i) \cdot \beta(i)$. In order to give the solution of the system with variable coefficients, we need a definition.

\begin{definition}
The symbol $A^{(k)}(n)$ is defined by $A^{(k)}(n) = A(n)A(n-1) \cdots A(n-(k-1))$, for $k = 1, 2, \ldots , n+1$. This definition can be extended to allow $k=0$ by letting $A^{(0)}(n) = I$, the identity matrix. (Note that $k$ represents the number of matrices being multiplied.)
\end{definition}

Using this definition, it can be shown \cite{Car08} that the solution of the system with variable coefficients is

\begin{equation} \label{eq:2Sec2}
f(n+1) = [A^{(n)}(t) \ast g(n)]_{t=n}.
\end{equation}

\noindent In the expression $[A^{(n)}(t) \ast g(n)]_{t=n}$, the exponent $n$ is the variable of the convolution, used to expand the expression $A^{(n)}(t) \ast g(n)$, while the argument $t$ stays fixed until after the convolution has been expanded, and then is substituted by $t=n$. That is, $[A^{(n)}(t) \ast g(n)]_{t=n} = [A^{(n)}(t) \cdot g(0)]_{t=n} + [A^{(n-1)}(t) \cdot g(1)]_{t=n} + \ldots + [A^{(1)}(t) \cdot g(n-1)]_{t=n} + [A^{(0)}(t) \cdot g(n)]_{t=n} = A(n) \cdots A(1)g(0) + A(n) \cdots A(2)g(1) + \ldots + A(n)g(n-1) + g(n)$.

Now suppose we have two $k$ by $k$ systems $f(n+1) = Af(n) + g(n)$ and $g(n+1) = Bg(n) + h(n)$, with constant coefficients. We can compose these two first-order systems to obtain a second-order system. The solution of this system gives $f$ in terms of $h$. It can be shown \cite{Car08} that its solution is

\begin{equation} \label{eq:2Sec3}
f(n+2) = A^n \ast [B^n \ast h(n)].
\end{equation}

\noindent Similarly, we could consider two $k$ by $k$ systems with variable coefficients: $f(n+1) = A(n)f(n) + g(n)$ and $g(n+1) = B(n)g(n) + h(n)$. The result of composing these two systems yields $f$ in terms of $h$. It can be shown \cite{Car08} that its solution is

\begin{equation} \label{eq:2Sec4}
f(n+2) = [A^{(n)}(t) \ast [B^{(n)}(s) \ast h(n)]_{s=n}]_{t=n+1}.
\end{equation}

\begin{example}
Let $f(n+1) = (n+1)f(n) + g(n)$ and $g(n+1) = ng(n) + h(n)$ be systems over $R=\mathbb{N}$, where $f(0) = 0$, $g(0) = 0$ and $h(n) = n+1$. Note that $A(n) = n+1$ and $B(n) = n$. Suppose we want to find $f(4)$. First, direct computation shows

\[ g(1) = 1, g(2) = 3, g(3) = 9. \]

\noindent And by using this, we obtain

\[ f(1) = 0, f(2) = 1, f(3) = 6, f(4) = 33. \]

\noindent We can check this result by using formula \eqref{eq:2Sec4}, which allows us to compute $f$ directly from $h$: \\

\begin{tabbing}
$f(4)$ \= $=$ \= $f(2+2) = [A^{(n)}(t) \ast [B^{(n)}(s) \ast h(n)]_{s=n}]_{t=n+1}|_{n=2}$ \\ \\
\> $= [A^{(2)}(t) [B^{(0)}(s) \ast h(0)]_{s=0} + A^{(1)}(t) [B^{(1)}(s) \ast h(1)]_{s=1} + A^{(0)}(t) [B^{(2)}(s) \ast h(2)]_{s=2}]_{t=3}$ \\ \\
\> $= [A^{(2)}(t)[B^{(0)}(0) h(0)] + A^{(1)}(t)[B^{(1)}(1) h(0) + B^{(0)}(1) h(1)] + A^{(0)}(t)[B^{(2)}(2) h(0) + B^{(1)}(2) h(1) + B^{(0)}(2) h(2)]]_{t=3}$ \\ \\
\> $= A^{(2)}(3)[B^{(0)}(0)h(0)] + A^{(1)}(3)[B^{(1)}(1)h(0) + B^{(0)}(1)h(1)] + A^{(0)}(3)[B^{(2)}(2)h(0) + B^{(1)}(2)h(1) + B^{(0)}(2)h(2)]$ \\ \\
\> $= 12[1] + 4[1 + 2] + 1[2 + 4 + 3] = 33$
\end{tabbing}
\end{example}

Formulas \eqref{eq:2Sec1} and \eqref{eq:2Sec2} will allow us to compute the $z$-transform of $f$ when $f$ is given by {\it a system} of linear recurrence equations. Formulas \eqref{eq:2Sec3} and \eqref{eq:2Sec4} will allow us to compute the $z$-transform of $f$ when $f$ is given by the composition of {\it two systems} of linear recurrence equations.

\section{The $z$-Transform of the Solution of a System}

The $z$-transform is a useful tool when working with functions defined as sequences, as in our case. We now define the $z$-transform.

\begin{definition}
Given a function \[ f = (f(0), f(1), \ldots , f(n), \ldots) \in R^{\mathbb{N}}, \] we define the $z$-transform of $f$ to be the series \[ f(0)/z^0 + f(1)/z^1 + \ldots + f(n)/z^n + \ldots = \sum_{n=0}^{\infty} f(n)/z^n. \] We denote the $z$-transform of $f$ by $z(f)$.
\end{definition}

Suppose now that $f$ is the $k$ by 1 vector of functions $f = \begin{pmatrix} f_1 \\ f_2 \\ \vdots \\ f_k \end{pmatrix}$, defined by $f = \begin{pmatrix} (f_1(0), f_1(1), \ldots , f_1(n), \ldots) \\ (f_2(0), f_2(1), \ldots , f_2(n), \ldots) \\ \vdots \\ (f_k(0), f_k(1), \ldots , f_k(n), \ldots) \end{pmatrix} \in R^{\mathbb{N}}_{k \times 1}$. We can apply the $z$-transform to every component, thus representing $f$ as $\begin{pmatrix} f_1(0)/z^0 + f_1(1)/z^1 + \ldots + f_1(n)/z^n + \ldots \\ f_2(0)/z^0 + f_2(1)/z^1 + \ldots + f_2(n)/z^n + \ldots \\ \vdots \\ f_k(0)/z^0 + f_k(1)/z^1 + \ldots + f_k(n)/z^n + \ldots \end{pmatrix} = \begin{pmatrix} \sum_{n=0}^{\infty} f_1(n)/z^n \\ \sum_{n=0}^{\infty} f_2(n)/z^n \\ \vdots \\ \sum_{n=0}^{\infty} f_k(n)/z^n \end{pmatrix}$. Alternatively, we can write $\begin{pmatrix} \sum_{n=0}^{\infty} f_1(n)/z^n \\ \sum_{n=0}^{\infty} f_2(n)/z^n \\ \vdots \\ \sum_{n=0}^{\infty} f_k(n)/z^n \end{pmatrix}$ as $\sum_{n=0}^{\infty} \begin{pmatrix} f_1(n) \\ f_2(n) \\ \vdots \\ f_k(n) \end{pmatrix} /z^n$, via an isomorphism. Note that this last expression is equal to $\sum_{n=0}^{\infty} f(n)/z^n$. Hence we can say that, in the general $k$ dimensional case, the image of $f \in R^{\mathbb{N}}_{k \times 1}$ under the $z$-transform is $\sum_{n=0}^{\infty} f(n)/z^n$.

In the constant case we are given a $k$ by $k$ system
\begin{equation*}
f(n+1) = Af(n) + g(n);
\end{equation*}
\noindent in the variable case we are given a $k$ by $k$ system
\begin{equation*}
f(n+1) = A(n)f(n) + g(n).
\end{equation*}
\noindent The solutions to these systems are, respectively,
\begin{equation*}
f(n+1) = A^n \ast g(n),
\end{equation*}
\noindent and
\begin{equation*}
f(n+1) = [A^{(n)}(t) \ast g(n)]_{t=n}.
\end{equation*}
\noindent These two expressions we have for $f(n+1)$ (one for the constant and one for the variable case) give the coefficients in the $z$-transform of $f$.

Since $f(0) = 0$, we obtain in both cases that $z(f) = \sum_{n=0}^{\infty} f(n)/z^n = f(0) + \sum_{n=0}^{\infty} f(n+1)/z^{n+1} = \sum_{n=0}^{\infty} f(n+1)/z^{n+1}$. In the constant case,
\begin{equation} \label{eq:3Sec1}
z(f) = \sum_{n=0}^{\infty} [A^n \ast g(n)] /z^{n+1},
\end{equation}
\noindent and in the variable case,
\begin{equation} \label{eq:3Sec2}
z(f) = \sum_{n=0}^{\infty} [A^{(n)}(t) \ast g(n)]_{t=n} /z^{n+1}.
\end{equation}

Using the definition of convolution, it is possible to rewrite the above series in a simpler fashion. In the constant case, \\

\begin{tabbing}
$z(f)$ \= $=$ \= $\sum_{n=0}^{\infty} [A^n \ast g(n)] /z^{n+1}$ \\ \\
\> $= [A^0 \ast g(0)] /z^1 + [A^1 \ast g(1)] /z^2 + [A^2 \ast g(2)] /z^3 + [A^3 \ast g(3)] /z^4 + \ldots$ \\ \\
\> $= [A^0 g(0)] /z^1$ \\
\> \> $+ [A^1 g(0) + A^0 g(1)] /z^2$ \\
\> \> $+ [A^2 g(0) + A^1 g(1) + A^0 g(2)] /z^3$ \\
\> \> $+ [A^3 g(0) + A^2 g(1) + A^1 g(2) + A^0 g(3)] /z^4 + \ldots$ \\ \\
\> $= [A^0 /z^0 + A^1 /z^1 + A^2 /z^2 + \ldots] g(0) /z^1$ \\
\> \> $+ [A^0 /z^0 + A^1 /z^1 + A^2 /z^2 + \ldots] g(1) /z^2$ \\
\> \> $+ [A^0 /z^0+ A^1 /z^1 + A^2 /z^2 + \ldots] g(2) /z^3 + \ldots$ \\ \\
\> $= \left[ \sum_{m=0}^{\infty} A^m /z^m \right] g(0) /z^1$ \\
\> \> $+ \left[ \sum_{m=0}^{\infty} A^m /z^m \right] g(1) /z^2$ \\
\> \> $+ \left[ \sum_{m=0}^{\infty} A^m /z^m \right] g(2) /z^3 + \ldots$ \\ \\
\> $= \sum_{n=0}^{\infty} \left[ \sum_{m=0}^{\infty} A^m /z^m \right] g(n) /z^{n+1}$ \\
\end{tabbing}

\noindent Notice that the series $\sum_{m=0}^{\infty} A^m /z^m$ is part of the coefficient of $1/z^{n+1}$, but it does not depend on $n$. We can rename this series as $S_A = \sum_{m=0}^{\infty} A^m /z^m$, and then write the series for $z(f)$, in terms of $g$, as $z(f) = \sum_{n=0}^{\infty} S_A g(n) /z^{n+1}$. This proves the following theorem.

\begin{theorem}
Let $f(n+1) = Af(n) + g(n)$ be a $k$ by $k$ system of nonhomogeneous linear recurrence equations with constant coefficients, with initial conditions equal to 0. Then the $z$-transform of $f$ is \begin{equation} z(f) = \sum_{n=0}^{\infty} S_A g(n) /z^{n+1}, \end{equation} \noindent where $S_A = \sum_{m=0}^{\infty} A^m /z^m$.
\end{theorem}

\noindent In the variable case, \\

\begin{tabbing}
$z(f)$ \= $=$ \= $\sum_{n=0}^{\infty} [A^{(n)}(t) \ast g(n)]_{t=n} /z^{n+1}$ \\ \\
\> $= [A^{(0)}(t) \ast g(0)]_{t=0} /z^1 + [A^{(1)}(t) \ast g(1)]_{t=1} /z^2 + [A^{(2)}(t) \ast g(2)]_{t=2} /z^3 + [A^{(3)}(t) \ast g(3)]_{t=3} /z^4 + \ldots$ \\ \\
\> $= [A^{(0)}(t) g(0)]_{t=0} /z^1$ \\
\> \> $+ [A^{(1)}(t) g(0) + A^{(0)}(t) g(1)]_{t=1} /z^2$ \\
\> \> $+ [A^{(2)}(t) g(0) + A^{(1)}(t) g(1) + A^{(0)}(t) g(2)]_{t=2} /z^3$ \\
\> \> $+ [A^{(3)}(t) g(0) + A^{(2)}(t) g(1) + A^{(1)}(t) g(2) + A^{(0)}(t) g(3)]_{t=3} /z^4 + \ldots$ \\ \\
\> $= [A^{(0)}(0) g(0)] /z^1$ \\
\> \> $+ [A^{(1)}(1) g(0) + A^{(0)}(1) g(1)] /z^2$ \\
\> \> $+ [A^{(2)}(2) g(0) + A^{(1)}(2) g(1) + A^{(0)}(2) g(2)] /z^3$ \\
\> \> $+ [A^{(3)}(3) g(0) + A^{(2)}(3) g(1) + A^{(1)}(3) g(2) + A^{(0)}(3) g(3)] /z^4 + \ldots$ \\ \\
\> $= [A^{(0)}(0) /z^0 + A^{(1)}(1) /z^1 + A^{(2)}(2) /z^2 + \ldots] g(0) /z^1$ \\
\> \> $+ [A^{(0)}(1) /z^0 + A^{(1)}(2) /z^1 + A^{(2)}(3) /z^2 + \ldots] g(1) /z^2$ \\
\> \> $+ [A^{(0)}(2) /z^0 + A^{(1)}(3) /z^1 + A^{(2)}(4) /z^2 + \ldots] g(2) /z^3 + \ldots$ \\ \\
\> $= \left[ \sum_{m=0}^{\infty} A^{(m)}(m) /z^m \right] g(0) /z^1$ \\
\> \> $+ \left[ \sum_{m=0}^{\infty} A^{(m)}(m+1) /z^m \right] g(1) /z^2$ \\
\> \> $+ \left[ \sum_{m=0}^{\infty} A^{(m)}(m+2) /z^m \right] g(2) /z^3 + \ldots$ \\ \\
\> $= \sum_{n=0}^{\infty} \left[ \sum_{m=0}^{\infty} A^{(m)}(m+n) /z^m \right] g(n) /z^{n+1}$ \\
\end{tabbing}

\noindent Note that the series $\sum_{m=0}^{\infty} A^{(m)}(m+n) /z^m$, which is part of the coefficient of $1/z^{n+1}$, is a function of $n$. By letting $S_A(n) = \sum_{m=0}^{\infty} A^{(m)}(m+n) /z^m$, we can write the series for $z(f)$, in terms of $g$, as $z(f) = \sum_{n=0}^{\infty} S_A(n) g(n) /z^{n+1}$. This shows the following theorem.

\begin{theorem}
Let $f(n+1) = A(n)f(n) + g(n)$ be a $k$ by $k$ system of nonhomogeneous linear recurrence equations with variable coefficients, with initial conditions equal to 0. Then the $z$-transform of $f$ is \begin{equation} z(f) = \sum_{n=0}^{\infty} S_A(n) g(n) /z^{n+1}, \end{equation} \noindent where $S_A(n) = \sum_{m=0}^{\infty} A^{(m)}(m+n) /z^m$.
\end{theorem}

\section{The $z$-Transform of the Solution of the Composition of two Systems}

Suppose $f$ is defined by
\begin{equation*}
f(n+1) = Af(n) + g(n), \ \mbox{where} \ g(n+1) = Bg(n) + h(n),
\end{equation*}
\noindent in the constant case;
\begin{equation*}
f(n+1) = A(n)f(n) + g(n), \ \mbox{where} \ g(n+1) = B(n)g(n) + h(n),
\end{equation*}
\noindent in the variable case. In both cases, we can solve for $f$ in terms of $h$. The solutions are
\begin{equation*}
f(n+2) = A^n \ast [B^n \ast h(n)],
\end{equation*}
\noindent in the constant case, and
\begin{equation*}
f(n+2) = [A^{(n)}(t) \ast [B^{(n)}(s) \ast h(n)]_{s=n}]_{t=n+1},
\end{equation*}
\noindent in the variable case. Again, notice that these expressions give the coefficients in the $z$-transform of $f$.

We assume that $f(0) = 0$ and $g(0) = 0$. Thus, $f(1) = 0$. We obtain in both cases that $z(f) = \sum_{n=0}^{\infty} f(n)/z^n$ = $f(0) + f(1) + \sum_{n=0}^{\infty} f(n+2)/z^{n+2}$ = $\sum_{n=0}^{\infty} f(n+2)/z^{n+2}$. In the constant case,
\begin{equation} \label{eq:4Sec1}
z(f) = \sum_{n=0}^{\infty} [A^n \ast [B^n \ast h(n)]] /z^{n+2},
\end{equation}
\noindent and in the variable case,
\begin{equation} \label{eq:4Sec2}
z(f) = \sum_{n=0}^{\infty} [A^{(n)}(t) \ast [B^{(n)}(s) \ast h(n)]_{s=n}]_{t=n+1} /z^{n+2}.
\end{equation}

Just as we did before, it is possible to write these series as simpler expressions. In the constant case, \\

\begin{tabbing}
$z(f)$ \= $=$ \= $\sum_{n=0}^{\infty} [A^n \ast [B^n \ast h(n)]] /z^{n+2}$ \\ \\
\> $= [A^0 \ast (B^0 \ast h(0))] /z^2 + [A^1 \ast (B^1 \ast h(1))] /z^3 + [A^2 \ast (B^2 \ast h(2))] /z^4 + \ldots$ \\ \\
\> $= [A^0(B^0 h(0))] /z^2$ \\
\> \> $+ [A^1(B^0 h(0)) + A^0(B^1 h(0)) + A^0(B^0 h(1))] /z^3$ \\
\> \> $+ [A^2(B^0 h(0)) + A^1(B^1 h(0)) + A^1(B^0 h(1)) + A^0(B^2 h(0)) + A^0(B^1 h(1)) + A^0(B^0 h(2))] /z^4 + \ldots$ \\ \\
\> $= A^0 B^0 h(0) /z^2$ \\
\> \> $+ A^1 B^0 h(0) /z^3 + A^0 B^1 h(0) /z^3 + A^0 B^0 h(1) /z^3$ \\
\> \> $+ A^2 B^0 h(0) /z^4 + A^1 B^1 h(0) /z^4 + A^1 B^0 h(1) /z^4 + A^0 B^2 h(0) z^4 + A^0 B^1 h(1) /z^4 + A^0 B^0 h(2) /z^4 + \ldots$ \\ \\
\> $= A^0 B^0 h(0) /z^2 + A^1 B^0 h(0) /z^3 + A^0 B^1 h(0) /z^3 + A^2 B^0 h(0) /z^4 + A^1 B^1 h(0) /z^4 + A^0 B^2 h(0) /z^4 + \ldots$ \\
\> \> $+ A^0 B^0 h(1) /z^3  + A^1 B^0 h(1) /z^4  + A^0 B^1 h(1) /z^4 + A^2 B^0 h(1) /z^5 + A^1 B^1 h(1) /z^5 + A^0 B^2 h(1) /z^5 + \ldots$ \\
\> \> $+ A^0 B^0 h(2) /z^4 + A^1 B^0 h(2) /z^5 + A^0 B^1 h(2) /z^5 + A^2 B^0 h(2) /z^6 + A^1 B^1 h(2) /z^6 + A^0 B^2 h(2) /z^6 + \ldots$ \\ \\
\> $= [A^0 B^0] h(0) /z^2 + [A^1 B^0 /z^1 + A^0 B^1 /z^1] h(0) /z^2 + [A^2 B^0 /z^2 + A^1 B^1 /z^2 + A^0 B^2 /z^2] h(0) /z^2 + \ldots$ \\
\> \> $+ [A^0 B^0] h(1) /z^3  + [A^1 B^0 /z^1  + A^0 B^1 /z^1] h(1) /z^3 + [A^2 B^0 /z^2 + A^1 B^1 /z^2 + A^0 B^2 /z^2] h(1) /z^3 + \ldots$ \\
\> \> $+ [A^0 B^0] h(2) /z^4 + [A^1 B^0 /z^1 + A^0 B^1 /z^1] h(2) /z^4 + [A^2 B^0 /z^2 + A^1 B^1 /z^2 + A^0 B^2 /z^2] h(2) /z^4 + \ldots$ \\ \\
\> $= [(A^0 \ast B^0) /z^0] h(0) /z^2 + [(A^1 \ast B^1) /z^1] h(0) /z^2 + [(A^2 \ast B^2) /z^2] h(0) /z^2 + \ldots$ \\
\> \> $+ [(A^0 \ast B^0) /z^0] h(1) /z^3 + [(A^1 \ast B^1) /z^1] h(1) /z^3 + [(A^2 \ast B^2) /z^2] h(1) /z^3 + \ldots$ \\
\> \> $+ [(A^0 \ast B^0) /z^0] h(2) /z^4 + [(A^1 \ast B^1) /z^1] h(2) /z^4 + [(A^2 \ast B^2) /z^2] h(2) /z^4 + \ldots$ \\ \\
\> $= \left[ \sum_{m=0}^{\infty} (A^m \ast B^m) /z^m \right] h(0) /z^2$ \\
\> \> $+ \left[ \sum_{m=0}^{\infty} (A^m \ast B^m) /z^m \right] h(1) /z^3$ \\
\> \> $+ \left[ \sum_{m=0}^{\infty} (A^m \ast B^m) /z^m \right] h(2) /z^4 + \ldots$ \\ \\
\> $= \sum_{n=0}^{\infty} \left[ \sum_{m=0}^{\infty} (A^m \ast B^m) /z^m \right] h(n) /z^{n+2}$ \\
\end{tabbing}

\noindent The series $\sum_{m=0}^{\infty} (A^m \ast B^m) /z^m$ is part of the coefficient of $1/z^{n+2}$ and it does not depend on $n$. Hence, we can let $S_{A,B} = \sum_{m=0}^{\infty} (A^m \ast B^m) /z^m$ and write the series for $z(f)$, in terms of $h$, as $z(f) = \sum_{n=0}^{\infty} S_{A,B} h(n) /z^{n+2}$. We have shown the following theorem.

\begin{theorem}
Let $f(n+1) = Af(n) + g(n)$ and $g(n+1) = Bf(n) + h(n)$ be two $k$ by $k$ systems of nonhomogeneous linear recurrence equations with constant coefficients, with initial conditions equal to 0. Then the $z$-transform of $f$ is \begin{equation} z(f) = \sum_{n=0}^{\infty} S_{A,B} h(n) /z^{n+2}, \end{equation} \noindent where $S_{A,B} = \sum_{m=0}^{\infty} (A^m \ast B^m) /z^m$.
\end{theorem}

\noindent In the variable case, \\

\begin{tabbing}
$z(f)$ \= $=$ \= $\sum_{n=0}^{\infty} [A^{(n)}(t) \ast [B^{(n)}(s) \ast h(n)]_{s=n}]_{t=n+1} /z^{n+2}$ \\ \\
\> $= [A^{(n)}(t) \ast [B^{(n)}(s) \ast h(n)]_{s=n}]_{t=n+1}|_{n=0} /z^2$ \\
\> \> $+ \!$ \= $\! [A^{(n)}(t) \ast [B^{(n)}(s) \ast h(n)]_{s=n}]_{t=n+1}|_{n=1} /z^3$ \\
\> \> $+ [A^{(n)}(t) \ast [B^{(n)}(s) \ast h(n)]_{s=n}]_{t=n+1}|_{n=2} /z^4 + \ldots$ \\ \\
\> $= [A^{(0)}(t) [B^{(0)}(s) \ast h(0)]_{s=0}]_{t=1} /z^2$ \\
\> \> $+ [A^{(1)}(t) [B^{(0)}(s) \ast h(0)]_{s=0} + A^{(0)}(t) [B^{(1)}(s) \ast h(1)]_{s=1}]_{t=2} /z^3$ \\
\> \> $+ [A^{(2)}(t) [B^{(0)}(s) \ast h(0)]_{s=0} + A^{(1)}(t) [B^{(1)}(s) \ast h(1)]_{s=1} + A^{(0)}(t) [B^{(2)}(s) \ast h(2)]_{s=2}]_{t=3} /z^4 + \ldots$ \\ \\
\> $= [A^{(0)}(t)[B^{(0)}(0) h(0)]]_{t=1} /z^2$ \\
\> \> $+ [A^{(1)}(t)[B^{(0)}(0) h(0)] + A^{(0)}(t)[B^{(1)}(1) h(0)] + A^{(0)}(t)[B^{(0)}(1) h(1)]]_{t=2} /z^3$ \\
\> \> $+ [A^{(2)}(t)[B^{(0)}(0) h(0)] + A^{(1)}(t)[B^{(1)}(1) h(0)] + A^{(1)}(t)[B^{(0)}(1) h(1)]$ \\
\> \> \> $+ A^{(0)}(t)[B^{(2)}(2) h(0)] + A^{(0)}(t)[B^{(1)}(2) h(1)] + A^{(0)}(t)[B^{(0)}(2) h(2)]]_{t=3} /z^4 + \ldots$ \\ \\
\> $= [A^{(0)}(1)[B^{(0)}(0) h(0)]] /z^2$ \\
\> \> $+ [A^{(1)}(2)[B^{(0)}(0) h(0)] + A^{(0)}(2)[B^{(1)}(1) h(0)] + A^{(0)}(2)[B^{(0)}(1) h(1)]] /z^3$ \\
\> \> $+ [A^{(2)}(3)[B^{(0)}(0) h(0)] + A^{(1)}(3)[B^{(1)}(1) h(0)] + A^{(1)}(3)[B^{(0)}(1) h(1)]$ \\
\> \> \> $+ A^{(0)}(3)[B^{(2)}(2) h(0)] + A^{(0)}(3)[B^{(1)}(2) h(1)] + A^{(0)}(3)[B^{(0)}(2) h(2)]] /z^4 + \ldots$ \\ \\
\> $= A^{(0)}(1)B^{(0)}(0) h(0) /z^2$ \\
\> \> $+ A^{(1)}(2)B^{(0)}(0) h(0) /z^3 + A^{(0)}(2)B^{(1)}(1) h(0) /z^3 + A^{(0)}(2)B^{(0)}(1) h(1) /z^3$ \\
\> \> $+ A^{(2)}(3)B^{(0)}(0) h(0) /z^4 + A^{(1)}(3)B^{(1)}(1) h(0) /z^4 + A^{(1)}(3)B^{(0)}(1) h(1) /z$ \\
\> \> \> $+ A^{(0)}(3)B^{(2)}(2) h(0) /z^4 + A^{(0)}(3)B^{(1)}(2) h(1) /z^4 + A^{(0)}(3)B^{(0)}(2) h(2) /z^4 + \ldots$ \\ \\
\> $= A^{(0)}(1)B^{(0)}(0) h(0) /z^2 + A^{(1)}(2)B^{(0)}(0) h(0) /z^3 + A^{(0)}(2)B^{(1)}(1) h(0) /z^3$ \\
\> \> $+ A^{(2)}(3)B^{(0)}(0) h(0) /z^4 + A^{(1)}(3)B^{(1)}(1) h(0) /z^4 + A^{(0)}(3)B^{(2)}(2) h(0) /z^4 + \ldots$ \\
\> \> $+ A^{(0)}(2)B^{(0)}(1) h(1) /z^3 + A^{(1)}(3)B^{(0)}(1) h(1) /z^4 + A^{(0)}(3)B^{(1)}(2) h(1) /z^4$ \\
\> \> $+ A^{(2)}(4)B^{(0)}(1) h(1) /z^5 + A^{(1)}(4)B^{(1)}(2) h(1) /z^5 + A^{(0)}(4)B^{(2)}(3) /z^2 h(1) /z^5 + \ldots$ \\
\> \> $+ A^{(0)}(3)B^{(0)}(2) h(2) /z^4 + A^{(1)}(4)B^{(0)}(2) h(2) /z^5 + A^{(0)}(4)B^{(1)}(3) h(2) /z^5$ \\
\> \> $+ A^{(2)}(5)B^{(0)}(2) h(2) /z^6 + A^{(1)}(5)B^{(1)}(3) h(2) /z^6 + A^{(0)}(5)B^{(2)}(4) h(2) /z^6 + \ldots$ \\ \\
\> $= [A^{(0)}(1)B^{(0)}(0) /z^0] h(0) /z^2 + [A^{(1)}(2)B^{(0)}(0) /z^1 + A^{(0)}(2)B^{(1)}(1) /z^1] h(0) /z^2$ \\
\> \> $+ [A^{(2)}(3)B^{(0)}(0) /z^2 + A^{(1)}(3)B^{(1)}(1) /z^2 + A^{(0)}(3)B^{(2)}(2) /z^2] h(0) /z^2 + \ldots$ \\
\> \> $+ [A^{(0)}(2)B^{(0)}(1) /z^0] h(1) /z^3 + [A^{(1)}(3)B^{(0)}(1) /z^1 + A^{(0)}(3)B^{(1)}(2) /z^1] h(1) /z^3$ \\
\> \> $+ [A^{(2)}(4)B^{(0)}(1) /z^2 + A^{(1)}(4)B^{(1)}(2) /z^2 + A^{(0)}(4)B^{(2)}(3) /z^2] h(1) /z^3 + \ldots$ \\
\> \> $+ [A^{(0)}(3)B^{(0)}(2) /z^0] h(2) /z^4 + [A^{(1)}(4)B^{(0)}(2) /z^1 + A^{(0)}(4)B^{(1)}(3) /z^1] h(2) /z^4$ \\
\> \> $+ [A^{(2)}(5)B^{(0)}(2) /z^2 + A^{(1)}(5)B^{(1)}(3) /z^2 + A^{(0)}(5)B^{(2)}(4) /z^2] h(2) /z^4 + \ldots$ \\ \\
\> $= [A^{(m)}(t) \ast B^{(m)}(m) /z^0]_{t=1}|_{m=0} h(0) /z^2 + [A^{(m)}(t) \ast B^{(m)}(m) /z^1]_{t=2}|_{m=1} h(0) /z^2$ \\
\> \> $+ [A^{(m)}(t) \ast B^{(m)}(m) /z^2]_{t=3}|_{m=2} h(0) /z^2 + \ldots$ \\
\> \> $+ [A^{(m)}(t+1) \ast B^{(m)}(m+1) /z^0]_{t=1}|_{m=0} h(1) /z^3 + [A^{(m)}(t+1) \ast B^{(m)}(m+1) /z^1]_{t=2}|_{m=1} h(1) /z^3$ \\
\> \> $+ [A^{(m)}(t+1) \ast B^{(m)}(m+1) /z^2]_{t=3}|_{m=2} h(1) /z^3 + \ldots$ \\
\> \> $+ [A^{(m)}(t+2) \ast B^{(m)}(m+2) /z^0]_{t=1}|_{m=0} h(2) /z^4 + [A^{(m)}(t+2) \ast B^{(m)}(m+2) /z^1]_{t=2}|_{m=1} h(2) /z^4$ \\
\> \> $+ [A^{(m)}(t+2) \ast B^{(m)}(m+2) /z^2]_{t=3}|_{m=2} h(2) /z^4 + \ldots$ \\ \\
\> $= \left[ \sum_{m=0}^{\infty} [A^{(m)}(t) \ast B^{(m)}(m)]_{t=m+1} /z^m \right] h(0) /z^2$ \\
\> \> $+ \left[ \sum_{m=0}^{\infty} [A^{(m)}(t+1) \ast B^{(m)}(m+1)]_{t=m+1} /z^m \right] h(1) /z^3$ \\
\> \> $+ \left[ \sum_{m=0}^{\infty} [A^{(m)}(t+2) \ast B^{(m)}(m+2)]_{t=m+1} /z^m \right] h(2) /z^4 + \ldots$ \\ \\
\> $= \sum_{n=0}^{\infty} \left[ \sum_{m=0}^{\infty} [A^{(m)}(t+n) \ast B^{(m)}(m+n)]_{t=m+1} /z^m \right] h(n) /z^{n+2}$ \\
\end{tabbing}

\noindent The series $\sum_{m=0}^{\infty} [A^{(m)}(t+n) \ast B^{(m)}(m+n)]_{t=m+1} /z^m$, which is part of the coefficient of $1/z^{n+2}$, is a function of $n$. We can let $S_{A,B}(n) = \sum_{m=0}^{\infty} [A^{(m)}(t+n) \ast B^{(m)}(m+n)]_{t=m+1} /z^m$, and thus write the series for $z(f)$, in terms of $h$, as $z(f) = \sum_{n=0}^{\infty} S_{A,B}(n) h(n) /z^{n+2}$. We have proved the following theorem.

\begin{theorem}
Let $f(n+1) = A(n)f(n) + g(n)$ and $g(n+1) = B(n)f(n) + h(n)$ be two $k$ by $k$ systems of nonhomogeneous linear recurrence equations with variable coefficients, with initial conditions equal to 0. Then the $z$-transform of $f$ is \begin{equation} z(f) = \sum_{n=0}^{\infty} S_{A,B}(n) h(n) /z^{n+2}, \end{equation}\noindent where $S_{A,B}(n) = \sum_{m=0}^{\infty} [A^{(m)}(t+n) \ast B^{(m)}(m+n)]_{t=m+1} /z^m$.
\end{theorem}

\bibliographystyle{plain}
\bibliography{MyBibliography}

\begin{thebibliography}{1}

\bibitem{Car08}
E.~Carta.
\newblock {\em Update Transducers and Linear Recurrence Equations over
  Semirings}.
\newblock PhD thesis, Cornell University, Ithaca, New York, 2008.

\bibitem{Dro05}
M.~Droste and P.~Gastin.
\newblock Weighted automata and weighted logics.
\newblock Technical report, Laboratoire de Sp\'{e}cification et
  V\'{e}rification, ENS de Cachan and CNRS, 2005.

\bibitem{Rue01}
J.~J. M.~M. Rutten.
\newblock Coinductive counting: Bisimulation in enumerative combinatorics
  (extended abstract).
\newblock Technical report, Centrum voor Wiskunde en Informatica, 2001.

\bibitem{Rut01}
J.~J. M.~M. Rutten.
\newblock Elements of stream calculus (an extensive exercise in coinduction).
\newblock Technical report, Centrum voor Wiskunde en Informatica, 2001.

\bibitem{Rue02}
J.~J. M.~M. Rutten.
\newblock Behavioral differential equations: a coinductive calculus of streams,
  automata, and power series.
\newblock Technical report, Centrum voor Wiskunde en Informatica, 2002.

\bibitem{Rut02}
J.~J. M.~M. Rutten.
\newblock Coinductive counting with weighted automata.
\newblock Technical report, Centrum voor Wiskunde en Informatica, 2002.

\bibitem{Sch62}
P.~Sch{\"{u}}tzenberger.
\newblock Finite counting automata.
\newblock {\em Information and Control}, 1962.

\end{thebibliography}

\end{document}